\documentclass[12pt]{JHEP3}
\usepackage{amsfonts,epsfig}

\def\be{\begin{equation}}
\def\ee{\end{equation}}
\def\bea{\begin{eqnarray}}
\def\eea{\end{eqnarray}}
\def\bei{\begin{itemize}}
\def\eei{\end{itemize}}
\def\bee{\begin{enumerate}}
\def\eee{\end{enumerate}}
\def\noN{\nonumber}
\def\ccr{\nonumber\\}
\def\eqa{\!&=&\!}

\def\lx{\left}
\def\rx{\right}
\def\la{\langle}
\def\ra{\rangle}
\def\pa{\partial}

\def\b{\beta}

\def\t{\tau}

\def\d{\delta}
\def\m{\mu}

\author{Fiorenzo~Bastianelli $^{a}$, Olindo~Corradini $^{a}$ and
  Pablo~A.~G.~Pisani $^{b}$\\\vskip2mm
$^{a}$ Dipartimento  di Fisica, Universit{\`a} di Bologna
and  INFN, Sezione di Bologna\\
Via Irnerio, 46 - Bologna I-40126, Italy
\\ \vskip2mm 
$^{b}$ IFLP, Departamento de F\'isica,
Facultad de Ciencias Exactas\\
UNLP, C.C. 67 (1900), La Plata, Argentina\\ \vskip1mm

E-mail:
\email{bastianelli@bo.infn.it}, \email{corradini@bo.infn.it}
\\ \hskip1.45cm\email{pisani@obelix.fisica.unlp.edu.ar}}

\abstract{
We study a worldline approach to quantum field theories on flat manifolds 
with boundaries. We consider the concrete case of a scalar field propagating 
on $\mathbb{R}_+ \times {\mathbb R}^{D-1}$ which leads us to study the 
associated heat kernel through a one dimensional (worldline) path integral.
To calculate the latter we map it onto an auxiliary path integral on the full 
${\mathbb R}^D$  using an image charge. The main technical difficulty lies in 
the fact that a smooth potential on  ${\mathbb R}_+ \times {\mathbb R}^{D-1}$ 
extends to a potential which generically fails to be smooth on ${\mathbb R}^D$.
This implies that standard perturbative methods fail and must be improved.
We propose a method to deal with this situation. As a result we recover 
the known heat kernel coefficients on a flat manifold with geodesic boundary, 
and compute two additional ones, $A_3$ and $A_\frac{7}{2}$.
The calculation becomes sensibly harder as the perturbative order increases,
and we are able to identify the complete $A_\frac{7}{2}$  with the help of 
a suitable toy model. Our findings show that the worldline approach is viable 
on manifolds with boundaries. Certainly, it would be desirable to improve our method
of implementing the worldline approach to further simplify the
perturbative calculations that arise in the presence of non-smooth potentials.}

\preprint{}

\title{Worldline approach to quantum field theories on 
flat manifolds with boundaries}

\begin{document}

\section{Introduction}

The worldline approach to quantum field theory (QFT) has been extensively
used to simplify calculations of many physical quantities like amplitudes, 
anomalies and effective actions, see \cite{Schubert:2001he} for a review. 
Recently this method has been extended to include the couplings
of particles of spin 0, 1/2 and 1 to external gravity
\cite{Bastianelli:2002fv,Bastianelli:2002qw,Bastianelli:2005vk}
together with some new applications \cite{Bastianelli:2004zp}.
Extensions to higher spin fields is also under study \cite{Bastianelli:2007pv}.
In this paper we investigate the possibility of using a worldline approach 
to quantum field theories defined on manifolds with boundaries. Previous
investigations of the worldline path integral approach on manifolds with
boundaries have already been carried out by using numerical Monte Carlo
simulations and successfully applied to Casimir energy
calculations~\cite{Gies:2003cv,Gies:2006cq}. 

QFTs on manifolds with boundaries have always attracted a certain interest, 
with applications ranging from Casimir energies to critical phenomena near 
boundaries, including worldsheet approaches to open strings.
More recently, they have drawn attention in the context of the Horava-Witten 
theory \cite{Horava:1995qa} and in the brane world scenarios,
as for example in \cite{Arkani-Hamed:2001is}.
Thus, it seems useful to investigate and develop new calculational tools
that may be used to study the properties of QFT with boundaries.

We investigate here an analytical approach based on first quantization, where a path 
integral over the coordinates of the field quanta is introduced. 
To reach directly the heart of the problem, we consider the case of 
a scalar field $\phi$ defined on a flat space with geodesic boundary, 
the $D$-dimensional manifold 
${\cal M}={\mathbb R}_+ \times {\mathbb R}^{D-1} $. 
We use euclidean conventions and consider a generic potential $U(\phi)$. 
The QFT action is given by
\be
S_{QFT}[\phi]= \int_{\cal M} d^Dx 
\Big ( \frac{1}{2}  \partial_\mu \phi \partial^\mu \phi + U(\phi) \Big )
\ee
and after specifying suitable boundary conditions (e.g. Dirichlet or Neumann)
one may compute the corresponding one-loop effective action,
which can be represented as
\bea
\Gamma[\phi] \eqa  - \log {\rm Det}^{-\frac{1}{2}} [ -\square +U''(\phi)] 
=  -{1\over 2} \int_0^\infty {dT\over T } \ 
{\rm Tr}\, e^{-T ( -\square +U''(\phi))} 
\ccr
\eqa
-{1\over 2} \int_0^\infty {dT\over T } \ \int_{PBC} 
{\cal D}x\ e^{-S[x]} \ .
\label{wl}
\eea
The path integral in the last line is evaluated 
with periodic boundary conditions (PBC) and contains the worldline action 
\bea
S[x] = \int_{0}^{T} 
d\tau \left ( 
{1\over 4 } \dot x^\mu \dot x_\mu +U''(\phi) \right ) 
\eea
which produces the differential operator $-\square +U''(\phi)$ as quantum 
hamiltonian. The scalar field $\phi(x)$ is defined on ${\cal M}$ and, as a 
consequence, the paths $x^\mu(\tau)$ are restricted to lie on ${\cal M}$,
see figure~\ref{fig1}.
Thus, one has to develop tools to calculate worldline path integrals
with such a restriction on the path integration variables.

\DOUBLEFIGURE{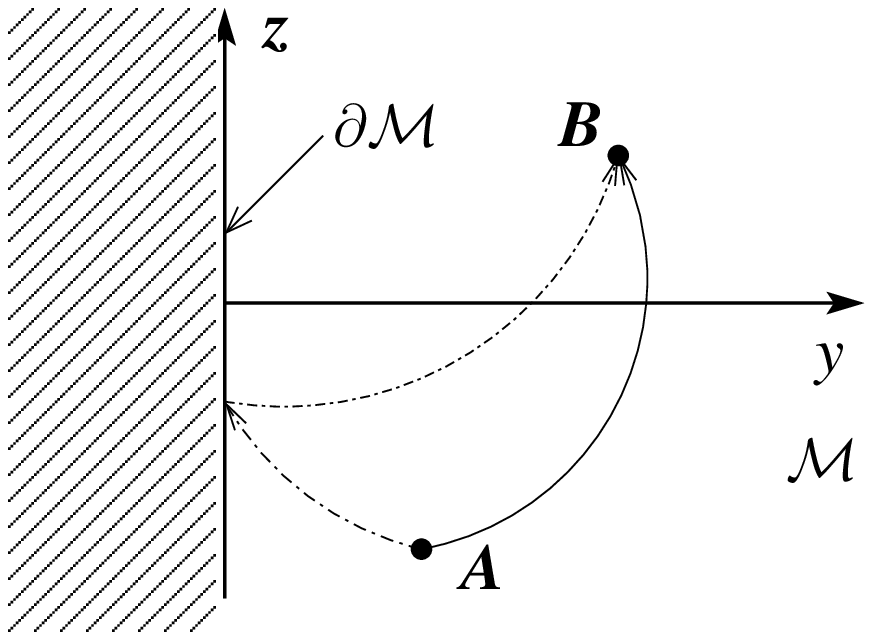,width=190pt}{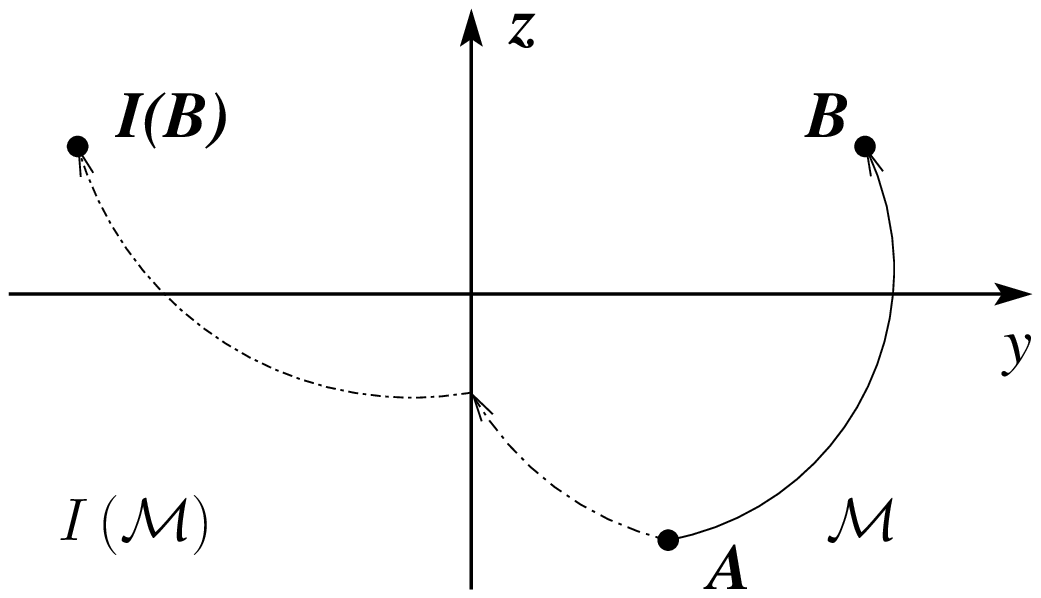,width=220pt}{Quantum paths from
 the initial point $A$ to the final point $B$, in the flat space ${\cal M}$
with boundary $\partial {\cal M}$.\label{fig1}}{The 
image charge  method: the bouncing path of figure 1 is mapped onto the path
 $A\rightarrow I(B)$.\label{fig2}}

General constructions of path integrals on target spaces with boundaries 
can be found in several textbooks, 
see for example \cite{Kleinert:2004ev,Chaichian:2001cz}.
However, the time slicing constructions presented there do not seem to 
assist in practical perturbative calculations.
One would like to have a reliable scheme to perform perturbative 
calculations in the presence of the boundary.
For this purpose we find it convenient to introduce an image charge and map 
the original problem to a new one defined on the full space 
${\mathbb R}^D$, see figure~\ref{fig2}.
The major difficulty now arises from the fact that
the potential extended on the full space
may be non-smooth at the fixed points of the ${\mathbb Z}_2$ identification 
that reintroduces the boundary. 
In this paper we develop a scheme to deal with this 
situation, and use it to compute perturbatively the matrix elements 
of the heat kernel $\la y|e^{-T ( -\square +U''(\phi))} |x\ra$ 
together with its trace, which appears in (\ref{wl}).
The non-smooth part of the interaction potential on  ${\mathbb R}^D$  
is going to be due to the presence of certain step functions.
Our scheme implements these step functions,
which contain the quantum field and sit inside correlation functions,
by transferring their effect to the limits of integration
of  suitably inserted completeness relations.

Several methods for computing the trace of the heat kernel on manifolds 
with boundaries have been studied and used in the literature, as for example 
in \cite{McKean:1967,Kennedy:1979ar,Branson:1990a,Cognola:1990kq,
McAvity:1990we,Branson:1995cm,Branson:1999jz}.
For a review and additional references one may consult 
\cite{Vassilevich:2003xt}. 
The method which is closer to the present one is presumably that developed 
by McAvity and Osborn \cite{McAvity:1990we}, where the DeWitt ansatz 
\cite{DW} for the heat kernel is generalized to the case with boundaries.
The novelty of our approach is that we use a path integral to calculate 
directly the heat kernel.
An advantage of the path integral method is that it is quite intuitive
and flexible, allowing to study the complete heat kernel 
and more general correlation functions, though we shall not investigate 
the latter here. 
Originally we also expected it to be  quite an efficient method, 
but we find that this expectation is not fully realized, 
and presumably it can be improved by 
devising alternative and more efficient
ways of calculating the perturbative expansion.
Nevertheless, in this paper we are able to reach higher perturbative orders for the trace 
of the heat kernel than those we could find in the literature 
\cite{Branson:1995cm,Vassilevich:2003xt},
namely $A_3$ and $A_\frac{7}{2}$, see eqs. (\ref{A3n}) and (\ref{A72}), 
but to identify the full $A_\frac{7}{2}$
we needed to use the result of a suitable toy model. 
In general, higher orders of the perturbative expansion 
produce multiple integrals
that are difficult to compute explicitly, even though they could 
be computed numerically, perhaps using the methods of 
refs.~\cite{Gies:2003cv,Gies:2006cq}. 
We use the toy model to identify precisely 
the value of one of these integrals.
 
After reviewing in section 2 the path integral approach to the heat kernel
on flat spaces without boundaries, we discuss in section 3 the case with 
a geodesic boundary, presenting first a brief description of our method 
and then its application to the half line and to 
$\mathbb{R}_+ \times {\mathbb R}^{D-1}$. 
We give the heat kernel coefficients $A_n$ with $n$ integer and half integer 
up to $n=\frac{7}{2}$.
We use a toy model to identify the value of a numerical  
coefficient entering the complete expression for $A_\frac{7}{2}$.
Finally we present our conclusions and an outlook in section 4.

\section{Heat kernel and path integrals}\label{two}

Let us start reviewing the standard definition of the heat kernel
and the way a path integral can be used to compute it.
This will also serve to define our notations.

Given the differential operator corresponding to
the hamiltonian of a nonrelativistic particle with unit mass
\be
H_x= -\frac{1}{2}\nabla^2_x + V(x)
\label{1}
\ee
where $\nabla^2_x$ is the laplacian in the cartesian coordinates $x^\mu$ 
of  flat space $\mathbb{R}^D$ and $V(x)$ an arbitrary potential, 
the heat kernel $K(x,y;\b)$
is the solution of the Schr\"odinger equation in euclidean time
(the heat equation)
with a particular boundary condition at time $\b=0$
\bea
&& -{\pa \over\pa\b} K(x,y;\b) = H_x K(x,y;\b)\ccr
&& K(x,y;0)= \d^D(x-y)~.
\eea
The heat kernel is formally given by the matrix element of the evolution
operator $e^{-\b  H}$,  
which in the Dirac bracket notation reads
\be
K(x,y;\b)= \la y|e^{-\b H}|x\ra  
\ee
and corresponds to the transition amplitude for the quantum system to start
at point $x^\mu$ and reach point $y^\mu$ in an euclidean time $\b$.
After setting $\beta=i t$ it gives the 
transition amplitude corresponding to the Schr\"odinger equation 
in real minkowskian time $t$, but we prefer to use  
the Wick rotated version with euclidean 
time\footnote{Substituting $\b \to 2T$ and $V(x) \to \frac{1}{2} U(\phi(x))$
one finds the heat kernel needed for the scalar field $\phi$
described in the introduction.}. 

Various methods for computing the heat kernel from
the Schr\"odinger equation with arbitrary potentials $V(x)$ 
have been developed in the literature. One of the most useful
is the path integral approach which produces $K(x,y;\b)$
as a sum over all paths $x^\m(t)$ linking the 
initial point $x^\m$ to the final point $y^\m$ in a time $\b$ 
\bea
K(x,y;\b)= 
\int_{x(0)=x}^{x(\b)=y} \hskip-0.5cm 
Dx\  e^{-S[x]}   
\eea
where the euclidean action $S[x]$ in the exponent 
is the one leading to the hamiltonian $H_x$, i.e. 
\bea
S[x] = \int_0^\b dt\, \bigg (
{1\over 2} \delta_{\mu\nu} \dot x^\mu \dot x^\nu + V(x)\bigg )~.
\eea
The symbol $Dx$ indicates the path integral measure
\be
Dx =\prod_{0<t<\b} d^D x(t)
\ee
where the range of $x^\m(t)$ is the full $\mathbb{R}^D$ for any $t$.

In general, 
it is not known how to compute the path integral for arbitrary potentials 
$V(x)$. A standard approximation method, useful 
in many contexts, is to compute  the path integral 
in a perturbative expansion for small propagation time $\b$
and small distances $\xi^\mu \equiv y^\m-x^\m$.
It is often useful to introduce a rescaled time $\t=t/\b$
and write the action as 
\bea
S[x] = \frac{1}{\b}\int_0^1 d\t\, \bigg (
{1\over 2} \delta_{\mu\nu} \dot x^\mu \dot x^\nu 
+ \beta^2 V(x)\bigg ) 
\label{8}
\eea
where the dots now refer to $\t$ derivatives.
Then one may split the action 
into a free part plus an interacting one
\bea
S[x] \eqa S_0[x] +S_{int}[x] \ccr
S_0[x] \eqa
\frac{1}{\b}\int_0^1 d\t\, 
{1\over 2} \delta_{\mu\nu} \dot x^\mu \dot x^\nu\ , \quad 
S_{int}[x]
=
\frac{1}{\b}\int_0^1 d\t\, \beta^2 V(x)~.
\eea
The path integral for the free part $S_0[x]$
is exactly calculable, and the interaction part can be treated as 
a perturbation. The terms generated by the interaction potential $V(x)$,
assumed to be smooth,
can be computed by Wick contracting the quantum fields 
and evaluating the emerging Feynman diagrams.
This method of computation gives an answer 
of the form
\bea
K(x,y;\b)= \frac{1}{(2\pi \b)^\frac{D}{2}}\, e^{-S_0[\bar x]}\, 
\Omega(x,y;\b)
\label{DWansatz1}
\eea
where $\bar x^\m(\t)=x^\m +(y^\m-x^\m)\t$ is the classical trajectory of 
the free equations of motion 
derived form $S_0[x]$, so that $S_0[\bar x]= \frac{1}{2\b} (x-y)^2$.
The term $(2\pi\b)^{-\frac{D}{2}}$
is the one loop correction which gives the correct normalization 
of the free path integral, and $\Omega(x,y;\b)$ contains the
perturbative corrections due to a nonvanishing potential $V$.
The perturbative calculation in terms of Feynman diagrams 
produces an answer of the form
\be
\Omega(x,y;\b)\sim\sum_{n=0}^\infty a_n (x,y) \b^n
\label{DWansatz2}
\ee
where $a_0(x,y)=1$ in flat space. The coefficients $a_n(x,y)$
are often called Seeley-DeWitt coefficients
and correspond to a $(n+1)$-loop calculation on the worldline\footnote{
This is generically true in curved space, as propagators go like $\beta$
and vertices like $\b^{-1}$, with $\b$ considered as the loop counting 
parameter. However in flat space the potential $V$ 
is the only source of vertices and contains an extra power
of $\beta^2$ which offsets this counting: each vertex with $V$
increases the loop counting by 2 because of the factor $\b^2$.}.
Thus we see that the DeWitt ansatz (\ref{DWansatz1}) -- (\ref{DWansatz2})
for solving the heat equation emerges naturally from the path integral.

For example, assuming a smooth potential $V$, one may compute perturbatively
to order $\b^\frac{7}{2}$ (counting $\xi^\mu = y^\mu-x^\mu \sim \sqrt\b$)
the transition amplitude
\bea
&& K(x,y;\beta) = { e^{- {\xi^2\over 2\b}} \over (2 \pi \b)^{D\over 2}}
\ccr 
&& \times \exp 
\bigg [ - \b  \Big (1 + \frac{1}{2} \xi\cdot \pa
+ \frac{1}{3!} (\xi\cdot \pa)^2
+ \frac{1}{4!} (\xi\cdot \pa)^3
+ \frac{1}{5!} (\xi\cdot \pa)^4
+ \frac{1}{6!} (\xi\cdot \pa)^5
\Big ) V(x) \ccr
&&  \hskip 1.2cm
-\frac{2 \b^2}{4!} 
\Big (1 + \frac{1}{2} \xi\cdot \pa
+ \frac{3}{20} (\xi\cdot \pa)^2
+ \frac{1}{30} (\xi\cdot \pa)^3
\Big ) \square V(x) \ccr
&& \hskip 1.2cm
+\frac{\b^3}{4!} 
\Big (1 + \frac{1}{2} \xi\cdot \pa \Big ) 
\Big(
(\pa_\mu V (x))^2 
-\frac{1}{10}  \square^2 V(x) \Big )+\cdots \bigg ] \ .
\label{eq:exp}
\eea
We have given the result in an exponentiated form since it is simpler 
to compute the connected worldline diagrams only.
Note that in this compact notation one considers $[\xi,\pa]=0$.
It is sometimes useful to present the result in a symmetrized form
as follows
\bea
&& K(x,y;\beta) 
 = 
{ e^{- {\xi^2\over 2\b}} \over (2 \pi \b)^{D\over 2}}
\exp \bigg [ \b \Big ( - \overline{V} +\frac{1}{12} 
\xi^\mu \xi^\nu \overline{ \pa_\mu \pa_\nu V} 
-\frac{1}{5!} \xi^\mu \xi^\nu \xi^\alpha \xi^\beta
\overline{ \pa_\mu \pa_\nu \pa_\alpha \pa_\beta  V} \Big )\ccr
&& + \frac{\beta^2}{5!}
\Big  ( -10 \overline{ \square V} + \xi^\mu \xi^\nu \overline{ 
\pa_\mu \pa_\nu \square V}\Big ) 
+ \frac{\beta^3}{5!} 
\Big 
( 5 \overline{ (\pa_\mu V)^2} 
- \frac{1}{2} \overline{ \square^2 V}\Big )
+\cdots   \bigg ]
\label{eq:whole-even}
\eea 
where we have introduced the notation $\overline{V}=\frac{1}{2}(V(x)+V(y))$, 
etc., so that the result is written in a more compact form
and the symmetry  $ x\leftrightarrow y$ is manifest.

Expanding the exponent one reads off the expansion in $\xi$ of the
Seeley-DeWitt coefficients $a_0, a_1,a_2, a_3$
\bea
a_0(x,y) \eqa 1 \ccr
a_1(x,y) \eqa 
 - \overline{V} +\frac{1}{12} \xi^\mu \xi^\nu \overline{ \pa_\mu \pa_\nu V} 
-\frac{1}{5!} \xi^\mu \xi^\nu \xi^\alpha \xi^\beta
\overline{ \pa_\mu \pa_\nu \pa_\alpha \pa_\beta  V}
+O(\xi^6) 
\ccr
a_2(x,y) \eqa  
 \frac{1}{2} \overline{V}^2-
 \frac{1}{5!}
\Big  ( 10 \overline{ \square V} - \xi^\mu \xi^\nu \overline{ 
\pa_\mu \pa_\nu \square V}\Big ) 
 -\frac{1}{12} 
\xi^\mu \xi^\nu \overline{V}\
\overline{ \pa_\mu \pa_\nu V}  
+O(\xi^4)
\ccr
a_3(x,y) \eqa  
- \frac{1}{3!} \overline{V}^3+
\frac{1}{4!} 
\Big 
( \overline{ (\pa_\mu V)^2} 
+2\overline{V}\ \overline{\square V}
- \frac{1}{10} \overline{ \square^2 V}
\Big ) +O(\xi^2)~. 
\eea
Their values at coinciding points $y^\mu= x^\mu$ (i.e. $\xi^\mu=0$) are 
easily obtained 
\bea
a_0(x,x)
\eqa 1 \ccr
a_1(x,x) \eqa - V \ccr
a_2(x,x) \eqa  {1 \over 2}  V^2 - {1\over 12} \square   V  \ccr
a_3(x,x) \eqa  -{1 \over 3!}  V^3 + {1\over 4!} \Big [
(\pa_\mu V)^2 + 2 V\square V -{1\over 10} \square^2   V 
\Big ] \ .
\eea
As for higher orders, 
one may find in \cite{Fliegner:1994zc} the heat kernel 
coefficients $a_n(x,x)$ (modulo total derivatives) 
up to $n=8$ included, a result which has been obtained 
using a worldline approach.

\section{Heat kernel and path integrals on a flat space with boundary}

\subsection{The method}
Now let us consider the case of a space with a boundary.
For definiteness we consider the flat space 
${\cal M}=\mathbb{R}_+\times \mathbb{R}^{D-1}$
with coordinates $x^\mu=(y,z^i)$ where 
$0 \leq y<\infty$ and $z^i \in \mathbb{R}^{D-1}$.
This space has a boundary $\partial {\cal M}=  \mathbb{R}^{D-1}$ located at
$y=0$, see figure~\ref{fig1}.
We wish to obtain the heat kernel for the operator
$H_x$ again as a path integral
\bea
K(x_1,x_2;\b)= 
\int_{x(0)=x_1}^{x(1)=x_2} \hskip-0.5cm  Dx\ e^{-S[x]}   
\eea
with the same action $S[x]$ as in (\ref{8}), but now summed over 
all paths $x^\m(\t)$ that lie  in ${\cal M}$.
In the presence of a boundary one must impose 
boundary conditions on $\partial {\cal M}$,
and for simplicity we assume either Dirichlet 
boundary conditions, $K(x,x_2;\b)= 0$ for $x\in \partial  {\cal M}$,
or Neumann boundary conditions,
$\partial_y K(x,x_2;\b)= 0$ for $x\in \partial  {\cal M}$
where $\partial_y $ is the derivative normal to the boundary.

It is evident that on such a manifold the free action $S_0$
has two local minima corresponding to the straight classical path 
$\bar x_1(\t)$ joining $x_1$ to $x_2$
and to the path $\bar x_2(\t)$ which bounces once at the boundary,
see figure~\ref{fig3}.
The path integral can be computed by summing over all fluctuations 
around these two classical paths. This produces an answer of the form
\bea
K(x_1,x_2;\b)= \frac{1}{(2\pi \b)^\frac{D}{2}}
\Big ( e^{-S_0[\bar x_1]}\, \Omega_1(x_1,x_2;\b) +\gamma\,  
e^{-S_0[\bar x_2]}\, \Omega_2(x_1,x_2;\b) \Big )
\label{MCAOansatz1}
\eea
where the $\Omega_i(x_1,x_2;\b)$ contain the perturbative loop corrections
around the classical paths $\bar x_i(\t)$, and $\gamma$ is a relative phase
between these paths which depends on the boundary conditions 
chosen on $\pa {\cal M}$.
This formula reproduces the generalized DeWitt ansatz introduced by
McAvity and Osborn to study the heat kernel on
spaces with boundaries~\cite{McAvity:1990we}. In the following we describe
a method for computing the path integral and obtain 
perturbatively the functions $\Omega_i(x_1,x_2;\b)$.
In particular we will focus on the calculations of the trace 
of the heat kernel
\be
{\rm Tr}\,  e^{-\b  H} = \int_{{\cal M}} d^Dx\,  K(x,x;\beta) =
\int_{PBC} Dx\ e^{-S[x]} 
\ee
where $PBC$ denote periodic boundary conditions for the trajectories 
on ${\cal M}$. 

\DOUBLEFIGURE{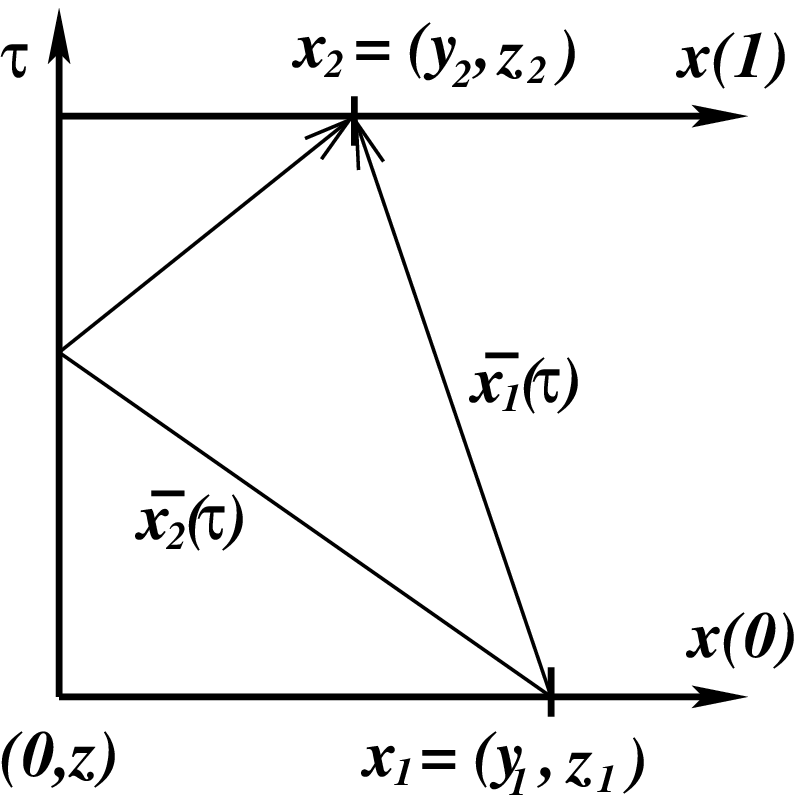,width=175pt}{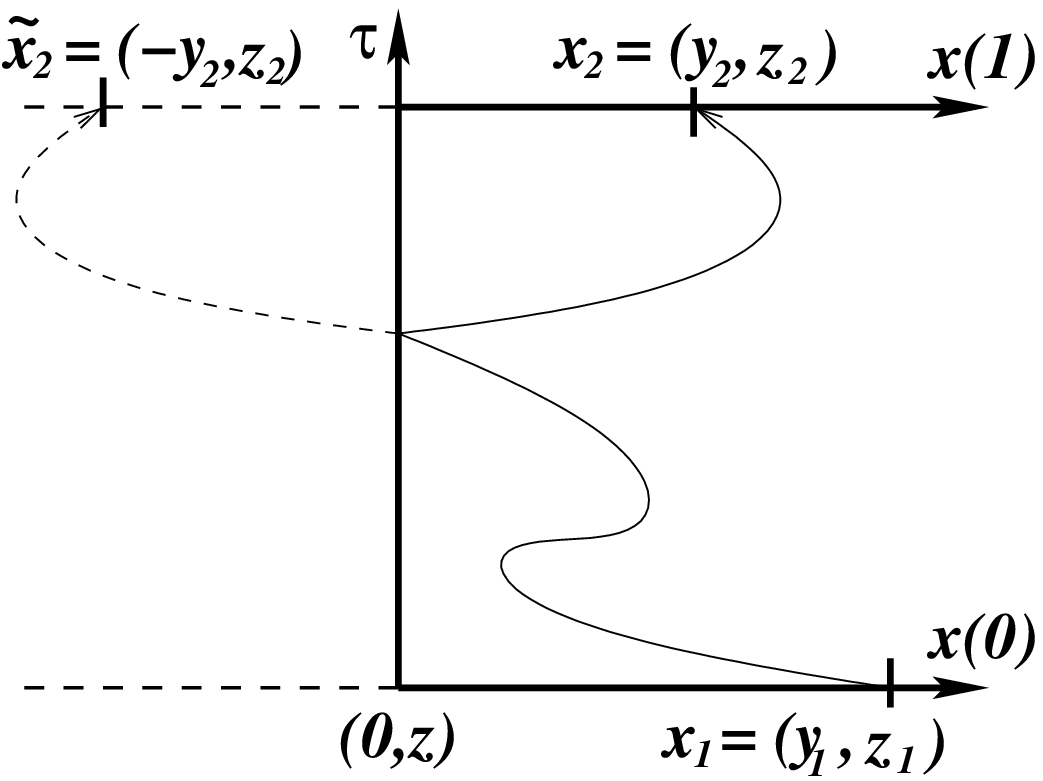,width=230pt}{Classical paths 
 in the flat space with boundary ${\cal M}$. 
On the vertical axis we have placed the time $\tau$.
\label{fig3}}{A bouncing quantum path and its image on 
$\mathbb{R}^D$.
\label{fig4}}

For evaluating the path integral over ${\cal M}$
it is convenient to use the method of images and double the space
to obtain the full $\mathbb{R}^D$.
Trajectories that bounce at the boundary are traded for trajectories 
that continue off the boundary with a reflected  path, see figure~\ref{fig4}.
A trajectory extended on the full $\mathbb{R}^D$ 
(dashed line in figure~\ref{fig4}) contributes the same weight  in the path 
integral as the original one (solid line), provided the potential is 
extended on $\mathbb{R}^D$ as an even function under the reflection $y \to -y$
\be
V(x) \to \tilde V(x)= \theta(y)V(y,z) +\theta(-y)V(-y,z)
\label{3.4}
\ee
where $\theta$ is the Heaviside (step) function.

Dirichlet or Neumann boundary conditions are special cases of the generic
ansatz (\ref{MCAOansatz1}): they correspond to setting $\gamma=-1$ or
$\gamma=+1$, respectively. 
Hence, the transition amplitude for such cases may be expressed as 
\be
K_{\cal M}(x_1,x_2;\b)= K_{\mathbb{R}^D}(x_1,x_2;\b)\mp 
K_{\mathbb{R}^D}(x_1,\tilde x_2;\b)~.
\label{eq:K}
\ee
The heuristic interpretation for $\gamma=-1$ is quite clear. In fact, notice that the 
second term in~(\ref{eq:K}) involves paths that cross the
boundary at least once. Hence, it is easy to convince oneself that 
-- due to the symmetry of the potential -- the only 
effect of such a term is to cancel all those paths in the 
first term that hit the boundary. 
In fact, setting $y_2=0$ ({\em all} paths now hit the boundary)  leads
to a vanishing kernel: therefore the wave function
evolved with the kernel
\bea
\psi(x_2,\beta) = \int_{\cal M} dx_1\, 
K_{\cal M}(x_1,x_2;\b)\,   
\psi(x_1,0)
\eea
satisfies Dirichlet boundary conditions.
For $\gamma=+1$ we know of no simple heuristic interpretation; however it is 
clear that differentiating (\ref{eq:K}) with respect to $y_2$ and setting 
$y_2=0$ leads to a vanishing result: the normal derivative of the wave 
function vanishes at the boundary. 
Hence, $\gamma=+1$ corresponds to Neumann boundary conditions.

The problem has thus been reduced to a path integral computation on a
 boundaryless manifold. However the path integral on $\mathbb{R}^D$ 
 with the potential $\tilde V$ is not easily computable by Taylor expanding 
the potential and using Wick contractions because of the step function 
$\theta(y)$ contained in $\tilde V$.
Even assuming that $V(x)$ is smooth enough to admit a Taylor expansion,
$\tilde V(x)$ in general is not.

Nevertheless, as a test, one could try to insert $\tilde V$ into (\ref{eq:exp}), which
was obtained by assuming smoothness of the potential. Evaluating the derivatives contained
in (\ref{eq:exp}) on $\tilde V$ would then produce delta functions 
and derivatives thereof, which use seems rather unwarranted and difficult to
implement consistently. 

Thus one has to study methods to compute perturbatively path integrals
that contain step functions of the quantum fields.
A method to address this problem is the following one.
Whenever a step function containing the quantum field $q(\tau)$ 
appears inside correlation function, such as $\la \theta(q(\tau_1)-a )\ra$, 
one may split the path integral into two parts using a completeness relation
at time $\tau_1$ 
(namely $ \mathbb{I} = \int_{-\infty}^\infty  dq_1 |q_1 \ra \la q_1 |$)
to transfer the constraint required by the step function into
the limit of integration  -- namely $\int_{a}^\infty  dq_1 $. 
Each of the two remaining parts are standard path integrals without 
step functions, but have boundary conditions linked by 
the $\int_{a}^\infty  dq_1 $ integration. 

Next we will exemplify this method to 
compute the heat kernel expansion for a particle on the half line.
This $D=1$ case is enough for our purposes,
since it contains the essential information.
At the end of the section we shall reconstruct  
the complete results for generic $D$.

\subsection{Particle on the half line}
We consider here the heat kernel on the half line ${\mathbb{R}_+}$
and denote by $x$ its single coordinate. For Dirichlet/Neumann
boundary conditions the expression for the transition amplitude reads
(in an obvious notation)
\bea
\la x_2|e^{-\beta  H}| x_1\ra_{\mathbb{R}_+} 
 =\la x_2, \beta| x_1, 0\ra_{\mathbb{R}_+} =
\la x_2, \beta| x_1, 
0\ra_{\mathbb{R}}\mp
\la \tilde x_2, \beta| x_1, 0\ra_{\mathbb{R}}
\label{eq:half-line}
\eea
where $\tilde x_2=-x_2$ is the coordinate of the image charge.
We now proceed to compute the path integral for 
the two terms on the right hand side, as explained earlier.

As usual in path integral computations one splits the quantum paths into a
classical path $x_{cl}(\t)$, satisfying the equations of motion, 
and quantum fluctuations $q(\t)$. 
We associate the boundary conditions with the classical path so that the 
quantum fluctuations $q(\t)$ have vanishing boundary conditions
\bea
&& x(\t) = x_{cl}(\t)+q(\t)= x_1 + (x_2-x_1)\t +q(\t)~,\\
&& q(0)=q(1)=0 ~.
\eea
We treat the potential $\tilde V$ as a perturbation for 
the free particle. Hence, the path integral for the full line, 
appearing in (\ref{eq:half-line}),  is given by
\bea\label{eq:whole-line}
\la x_2, \beta| x_1, 0\ra_{\mathbb{R}} &=& 
e^{-S_0[x_{cl}]} \int_{q(0)=0}^{q(1)=0} \hskip-0.5cm Dq\ 
e^{-S_0[q]}\exp\left(-\beta\int_0^1d\t\ \tilde
V(x_{cl}(\t)+q(\t))\right)\\ 
&=&  e^{-{1\over 2\beta}(x_2-x_1)^2} \int_{q(0)=0}^{q(1)=0} \hskip-0.5cm Dq\ 
e^{-S_0[q]}
\Biggl(1-\beta\int_0^1d\t_1\ \tilde V(x_{cl}(\t_1)+q(\t_1))\noN\\
&+&{\beta^2\over 2!}\int_0^1d\t_1\int_0^1d\t_2\ \tilde
V(x_{cl}(\t_1)+q(\t_1))\,
\tilde V(x_{cl}(\t_2)+q(\t_2))+O(\tilde V^3)\Biggr)\noN
\eea
where $S_0$ is the action for the free particle, $\tilde V=0$. 
The first term in the expansion simply yields the path integral normalization
\bea
\int_{q(0)=0}^{q(1)=0} \hskip-0.5cm Dq\ 
e^{-S_0[q]} = {1\over (2\pi\beta)^{1\over 2}}~,
\noN
\eea
whereas the rest contributes to the perturbative corrections to
the heat kernel. 

\DOUBLEFIGURE{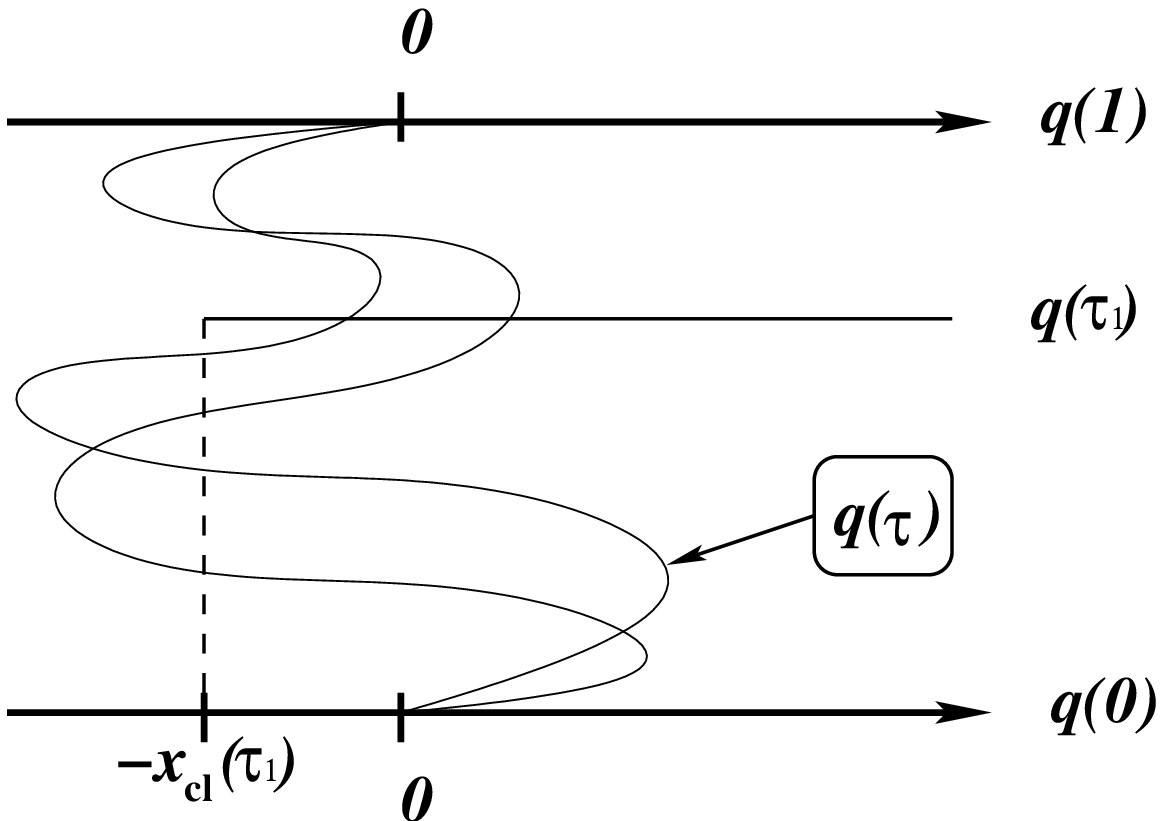,width=210pt}{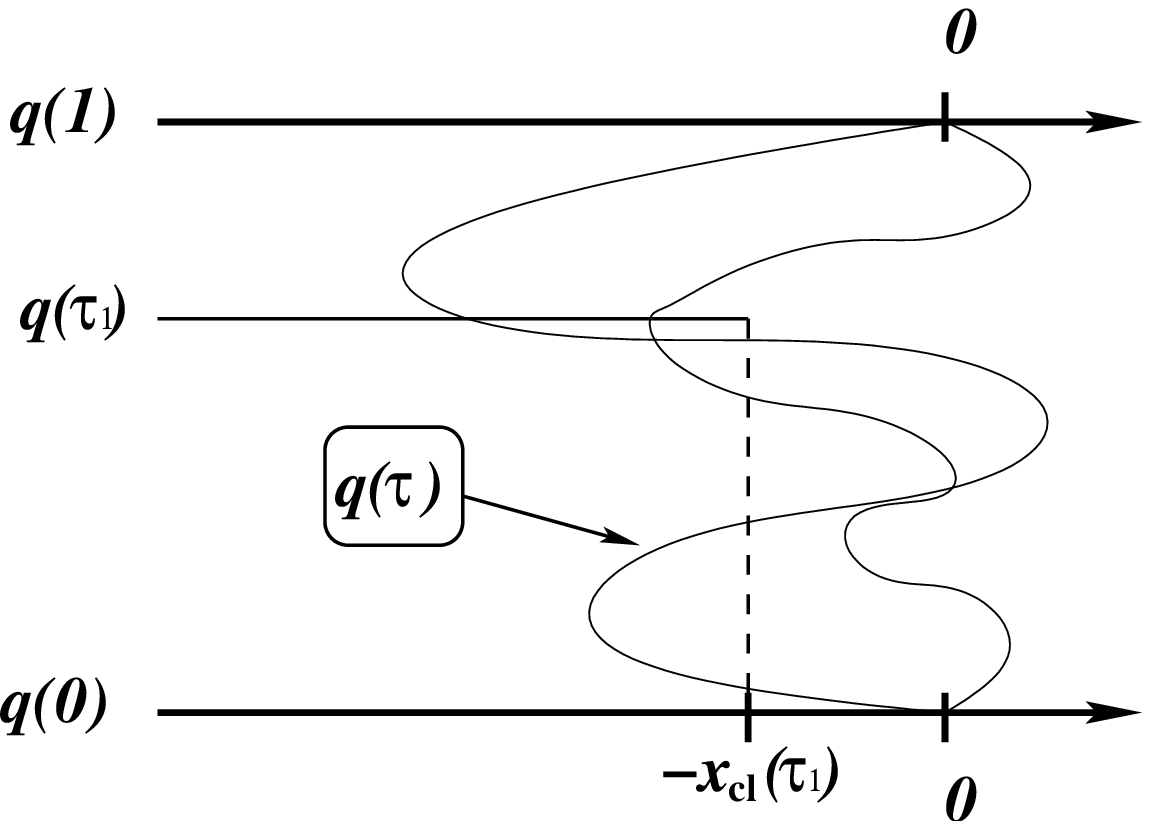,width=205pt}{Quantum paths 
constrained by  $\theta(x_{cl}(\tau_1)+q(\tau_1))$.
The time $\tau$ runs along the vertical axis.
\label{fig5}}{Quantum paths 
 constrained by  $\theta(-x_{cl}(\tau_1)-q(\tau_1))$.\label{fig6}}

We have managed to compute the short-time perturbative expansion of the
transition amplitude up to order $\beta^{7\over 2}$, 
 considering $x_1-x_2\sim \sqrt{\beta}$.
In the following we only give a detailed description of the contributions 
coming from a single insertion of the perturbation $V$, 
since this is enough  to exemplify the method discussed previously.
In principle this method can be applied to all orders in $V$.

To compute the aforementioned contributions we first extract the 
integral over the worldline (namely $\int_0^1d\t_1 $)
out of the path integral; we will perform it at 
the end. Then, using (\ref{3.4}), we have
\bea
 &&\int_{q(0)=0}^{q(1)=0} \hskip-0.5cm Dq\ 
e^{-S_0[q]}
\biggl[\theta(x_{cl}(\tau_1)+q(\tau_1)) V(x_{cl}(\tau_1)+q(\tau_1))\nonumber\\
&&\hskip3cm +\theta(-x_{cl}(\tau_1)-q(\tau_1))
V(-x_{cl}(\tau_1)-q(\tau_1)) \biggr]~.
\label{V-vev}
\eea
Note that the
constraints for the quantum field $q(\t)$, associated with the Heaviside
functions, are localized at time $\t=\tau_1$. Each of the two terms
in (\ref{V-vev}) can thus be manipulated by splitting the path integral into two
parts, as follows. In the first term the constraint acts as depicted
in figure \ref{fig5}: we can thus compute the path integral as a convolution
between a free path integral with boundary conditions $q(0)=0,\ q(\tau_1)=y$ 
and a free path integral with boundary conditions $q(\tau_1)=y,\ q(1)=0$, 
where $y$ is integrated over and satisfies the constraint
$y\geq -x_{cl}(\tau_1)$ because of the step function.
We thus have    
\bea
 &&\int_{q(0)=0}^{q(1)=0} \hskip-0.5cm 
Dq\ e^{-S_0[q]}\,
\theta(x_{cl}(\tau_1)+q(\tau_1))\ V(x_{cl}(\tau_1)+q(\tau_1))\nonumber\\&&= 
 \int_{-x_{cl}(\tau_1)}^\infty \hskip-0.5cm dy\
\left (
\int_{q(0)=0}^{q(\tau_1)=y} \hskip-0.5cm Dq\ e^{-S_0[q]} \right )  
V(x_{cl}(\tau_1)+y)
\left ( \int_{q(\tau_1)=y}^{q(1)=0} \hskip-0.5cm Dq\ e^{-S_0[q]} \right )
\nonumber\\&&= 
 \int_{-x_{cl}(\tau_1)}^\infty \hskip-0.5cm dy\ 
{e^{-{1\over 2\beta\tau_1(1-\tau_1)}y^2}
\over 2\pi\beta\sqrt{\tau_1(1-\tau_1)}}\ V(x_{cl}(\tau_1)+y) \ .
\label{theta-one}
\eea

In a similar fashion, in the second term of~(\ref{V-vev}) the
constraint acts as depicted in figure~\ref{fig6}, and the path integral reads
\bea
&&\int_{q(0)=0}^{q(1)=0} \hskip-0.5cm Dq\ 
e^{-S_0[q]}\
\theta(-x_{cl}(\tau_1)-q(\tau_1))\ V(-x_{cl}(\tau_1)-q(\tau_1))\nonumber\\&&=
 \int^{-x_{cl}(\tau_1)}_{-\infty} \hskip-0.5cm dy\ 
{e^{-{1\over 2\beta\tau_1(1-\tau_1)}y^2}
\over 2\pi\beta\sqrt{\tau_1(1-\tau_1)}}\ V(-x_{cl}(\tau_1)-y)~.
\label{theta-two}
\eea
Hence, the complete contribution of a single $V$ insertion
to the whole line path integral can be written as
\bea
&&\beta\, {e^{-{1\over 2\beta}(x_2-x_1)^2}\over (2\pi\beta)^{1\over 2}}
\int_0^1d\t {1\over \sqrt{2\pi\beta\t(1-\t)}}\ \Biggl(
-
\int_{-\infty}^{+\infty}\hskip-0.25cm dy\ e^{-{y^2\over 2\b\t(1-\t)}}\
V(x_{cl}(\t)+y)\ccr
&&+
\int_{-\infty}^{-x_{cl}(\t)}\hskip-0.5cm dy\ e^{-{y^2\over 2\b\t(1-\t)}}
\biggl[V(x_{cl}(\t)+y)-V(-x_{cl}(\t)-y)\biggr]\Biggr)~,\label{eq:whole}
\eea
where the constraint only sits in the integration limit 
of the second term. Higher order contributions in $V$ can 
analogously be split into one unconstrained part where 
the ``quantum field'' $y=q(\t)$ may run over the whole real axis, 
and parts that depend upon the constraints: the latter are 
parity-odd. A few observations are in order.\\
$\bullet$ For an even potential $V$ the parity-odd terms vanish
identically. In such a case it is possible to write the heat kernel 
as a power series in integer powers of $\beta$ and  $x_2-x_1$. In
fact, upon Taylor expanding the potential about the initial 
point $x_1$, the unconstrained gaussian integral over $y$ singles out 
only integer powers of $\beta$,
and one reproduces the expression~(\ref{eq:whole-even}),
with $x=x_1,\ y=x_2$ and $\xi= x_2-x_1$,
plus an image charge contribution.
We can thus write down the final
expression for the heat kernel on the half line
\bea\label{eq:half-even}
\la x_2, \beta| x_1, 0\ra_{\mathbb{R}_+}  
&=&{ e^{-{\xi^2\over 2\beta}}\over (2\pi\beta)^{1\over 2}}
\exp\Biggl[-\beta \overline{V}-{\beta^2\over 2\cdot 3!}\overline{\partial^2 V}
\left(1-\xi^2/
    \beta\rx)\ccr
&&-{\beta^3\over 2\cdot 5!}\overline{\partial^4 V}\left(1-2\xi^2/
    \beta+2\xi^4/\beta^2\rx)+{\beta^3\over
      4!}{\overline{\partial V}}^2+O(\beta^4)\Biggr]\ccr[2mm] 
&\mp&{ e^{-{\tilde \xi^2\over 2\beta}}\over (2\pi\beta)^{1\over 2}}
\exp\Biggl[-\beta \overline{V}-{\beta^2\over 2\cdot
    3!}\overline{\partial^2 V}\left(1-\tilde \xi^2/
    \beta\rx)\ccr&&-{\beta^3\over 2\cdot 5!}\overline{\partial^4 V}
\left(1-2\tilde \xi^2/
    \beta+2\tilde \xi^4/\beta^2\rx)+{\beta^3\over
      4!}{\overline{\partial V}}^2+O(\beta^4)\Biggr]\eea 
where $\tilde \xi=-x_2-x_1$. 
The expansion above clearly corresponds to the McAvity-Osborn ansatz
(\ref{MCAOansatz1}), with $\Omega_i$'s both expressed as 
integer power series in $\beta$. 
In the present case with even potentials, 
it is clear that the heat kernel is also correctly reproduced  
by the conventional perturbative calculation of the path integral
through the Wick's theorem, as discussed in section \ref{two}.\\
$\bullet$ For a generic potential $V$, the constraint-independent part of the 
expansion yields again 
the expression (\ref{eq:half-even}), except that now $\tilde{\overline{V}}=
{1\over 2} (V(-x_2)+V(x_1))$ must appear in the second exponential in place of
$\overline{V}={1\over 2} (V(x_2)+V(x_1))$, and so on.
However, the constraint-dependent part now is not
vanishing and does not seem to be naively expressible as a power series in
$\beta$ and $x_2-x_1$, as the latter 
appears nontrivially in the integration limits. This complication
corresponds to the difficulty of giving a
reliable interpretation of the Taylor expansion of the Heaviside function,
pointed out in the previous section.\\ 
$\bullet$ The method described above allows to compute the perturbative
expansion in $\beta$ of the partition function
\bea
{\rm Tr}\ e^{-\beta \hat H} &=& \int_0^\infty dx\ \la x, \beta| x,
0\ra_{\mathbb{R}_+} 
\nonumber\\ &=&
\int_0^\infty dx\ \la x, \beta| x, 0\ra_{\mathbb{R}}\mp
\int_0^\infty dx\ \la -x, \beta| x, 0\ra_{\mathbb{R}}
\label{part-fn}
\eea
for which we need to specialize expression (\ref{eq:whole}), and 
analogous higher-order contributions, to\\[2mm]
(i) $x_1=x_2 =x\quad\Longrightarrow\quad x_{cl}(\t) = x$\\ [1mm]
  and\\ [1mm]
(ii) $x_1=-x_2 =x\quad\Longrightarrow\quad x_{cl}(\t) = x(1-2\t)$~.\\[2mm]
\noindent Thus we can proceed to compute the small time expansion of the 
 partition function up to order $\beta^{7\over 2}$:\\[2mm]
(i) In this case, the constraint-independent parts of the potential
 insertions, such as the first term of (\ref{eq:whole}), yield bulk
 contributions, whereas the constraint-dependent parts, after Taylor expanding
 the potential about the boundary, single out boundary terms. Hence, after
 some tedious algebra, we have 
\bea
\int_0^\infty dx\ \la x, \beta| x, 0\ra_{\mathbb{R}}
&=& {1\over (2\pi\beta)^{1\over 2}} \Biggl[\int_0^\infty dx \ \exp\Biggl(-\beta
  V(x)-\beta^2{\partial^2 V(x)\over 2\cdot 3!} 
+\beta^3{(\partial V(x))^2\over 4!}\noN
\\&-& \beta^3{\partial^4 V(x)\over 2\cdot  5!}
\Biggr)
-\beta^2{\partial V(0)\over
    2\cdot3!}
+\beta^3\lx({ V\partial V(0)\over 2\cdot3!} 
-{ \partial^3 V(0)\over
  2\cdot5!}\rx)
\ccr&+&\sqrt{\pi\over 8}\,\beta^{7\over 2}\, {(\partial V(0))^2\over  2^6}  
+O(\beta^4)\Biggr]~.\label{3.17}
\eea\\[-1mm]
(ii) For such a case, the presence of the non trivial classical action
 $S_0[x_{cl}]=2x^2/\beta$  allows for a Taylor expansion of the potential about the boundary
 location: the integral over the manifolds is convergent and gives
\bea
\int_0^\infty dx\ \la -x, \beta| x, 0\ra_{\mathbb{R}}
&=&{1\over (2\pi\beta)^{1\over 2}}\Biggl\{\sqrt{\pi\beta\over 8}\Biggl[1-\beta
V(0)+{\beta^2\over 2}\lx(V^2(0)-{\partial^2 V(0)\over 4}
\rx)
\noN
\\&+&
\beta^3\lx(-{V^3(0)\over 3!}+{3(\partial V(0))^2\over 2^4} 
-{\partial^4 V(0)\over 2^7}\rx)
 \Biggr]-\beta^2{\partial V(0)\over 4} 
\ccr&+&\beta^3\lx({V\partial V(0)\over 4}-{\partial^3 V(0)\over
   2\cdot4!}\rx)+O(\beta^4)\Biggr\} \ .
\label{3.18}
\eea
Combining (\ref{3.17}) and (\ref{3.18}) we get the final result 
\bea
{\rm Tr}\ e^{-\beta \hat H} &=& \int_0^\infty dx\ \la x, \beta| x,
0\ra_{\mathbb{R}_+} 
\\ &=&
{1\over (2\pi\beta)^{1\over 2}} \Biggl[\int_0^\infty dx \ \exp\Biggl(-\beta
  V(x)-\beta^2{\partial^2 V(x)\over 2\cdot 3!}+\beta^3{(\partial V(x))^2\over 4!}\nonumber\\
&&-\beta^3{\partial^4 V(x)\over 2\cdot
  5!}\Biggr)
+\sqrt{\pi\over 8}\,\beta^{7\over 2}\, {(\partial V(0))^2\over 2^6}  
\ccr[2mm]
&\mp &\sqrt{\pi\beta\over 8}\Biggl[1-\beta
V(0)+{\beta^2\over 2}\lx(V^2(0)-{\partial^2 V(0)\over 4}\rx)\ccr
&&+\beta^3\lx(-{V^3(0)\over 3!}
+{V\partial^2 V(0)\over 2^3}+
{(\partial V(0))^2\over
  4!}-{\partial^4 V(0)\over 2^7}\rx) \Biggr]\ccr[2mm]
&+&\beta^2
\left\{\begin{array}{c}
1\\
-2
\end{array}\rx\}
{\partial V(0)\over 3!} +\beta^3\lx(\left\{\begin{array}{c}
-1\\
2
\end{array}\rx\}{V\partial V(0)\over 3!}+
\left\{\begin{array}{c}
2\\
-3
\end{array}\rx\}
{\partial^3 V(0)\over5!}\rx)+O(\beta^4)\Biggr]\ . \noN
\eea
This is the expansion of the partition function
with Dirichlet/Neumann boundary conditions, valid up to order 
$\beta^{7\over 2}$, where the upper coefficient in the symbol $\{ : \}$
 refers to Dirichlet boundary conditions, and the lower one to 
Neumann boundary conditions. At this point
the integrated heat kernel coefficients for the half line can be read off 
immediately. 
However, it is not difficult to reconstruct
such coefficients for arbitrary $D$.
Defining 
 \bea
{\rm Tr}\ e^{-\beta H} \sim 
{1\over (2\pi\beta)^{D\over 2}}\sum_{n\in {\mathbb N}/2} A_n \beta^n~.
\eea
we obtain
\bea
A_0 &=& \int_{\cal M} 1\label{A0}\\[3mm]
A_{1\over 2} &=& \mp \sqrt{\pi\over 8}\int_{\partial\cal M} 1\\[3mm]
A_1 &=& -\int_{\cal M} V\label{A1}\\[3mm]
A_{3\over 2} &=& \pm \sqrt{\pi\over 8} 
\int_{\partial\cal M} V\label{A32}\\[3mm]
A_2 &=& \int_{\cal M} \Biggl({V^2\over
    2}-{\Box V\over 12}\Biggr)+ 
\left\{\begin{array}{c}
1\\
-2
\end{array}\rx\}\int_{\partial\cal M}
{\partial_y V\over 3!}\\[3mm]
A_{5\over 2} &=& \mp \sqrt{\pi\over 8}
\int_{\partial\cal M}\Biggl[{V^2\over 2}-\lx({\partial_y^2 \over
    8}+{\partial_i^2 \over 12}\rx)V\Biggr] \label{A52}\\[3mm]
A_3 &=& \int_{\cal M} \Biggl(-{V^3\over
    3!}+{(\partial_\mu V)^2\over 4!}+{V \Box V\over 2\cdot 3!}-{\Box^2 V\over 2\cdot
  5!}\Biggr)\noN\\&+&\int_{\partial\cal M}
\Biggl[\left\{\begin{array}{c}
-1\\
2
\end{array}\rx\}{V\partial_y V\over 3!}+
\left\{\begin{array}{c}
2\\
-3
\end{array}\rx\}
{\partial_y^3 V\over5!}+
\left\{\begin{array}{c}
3\\
-7
\end{array}\rx\}
{\partial_y\partial_i^2 V\over2\cdot 5!}\Biggr] \label{A3n}
\\[3mm]
A_{7\over 2} &=& \sqrt{\pi\over 8} \int_{\partial\cal M}
\Biggl[\mp\Biggl[-{V^3\over 3!}+{(\partial_i V)^2\over 4!}
+V\lx({\partial_y^2\over 2^3}+{\partial_i^2\over 12}\rx)V
 \ccr
&&\hskip2cm
-\lx({\partial_y^4 \over 2^7}+
{\partial_y^2\partial_i^2\over 4\cdot 4!}+{\partial_i^4\over 10\cdot 4!}\rx)V
\Biggr]+\left\{\begin{array}{c}
-5 \\ 7
\end{array}\rx\} {(\partial_y V)^2 \over 2^6}\Biggr]\label{A72} \ .
\eea
Before concluding this subsection let us mention some caveats of our
method. We have found it increasingly harder to 
calculate explicitly the numerical value of some terms.
On the other hand our method allows to write 
all such numerical values in terms of multiple integrals of finite functions over  
compact domains which could eventually be calculated numerically. 
Since in the present manuscript we only 
encountered difficulties for one numerical coefficient in $A_{7\over 2}$, and
precisely the
one related to $(\partial_y V)^2$,  we decided to fix it by
using a toy model, which is discussed next.

\subsection{Toy model: $V(x)=a\,x$}

The asymptotic expansion of $e^{-\beta H}$ for small
$\beta$ can also be obtained from the large-$\lambda$ asymptotic 
expansion of the resolvent $(H+\lambda)^{-1}$.
Indeed, since the trace of the resolvent is the Laplace transform
of the heat kernel trace, it admits the following representation for large
$\lambda$ \cite{Vassilevich:2003xt} 
\begin{equation}\label{asyres}
    {\rm Tr}\,(H+\lambda)^{-1}\sim
    \frac{1}{\sqrt{2\pi}}\ \sum_{n\in\mathbb{N}/2}
    \Gamma\left(n+\frac{1}{2}\right)\,A_{n}\,
    \lambda^{-n-\frac{1}{2}}~.
\end{equation}
We use such an expansion to compute the $(\partial_y V)^2$ term in the coefficient
$A_{7\over 2}$,  
considering a toy model that can be explicitly solved in the
half line, namely the
one-dimensional Schr\"odinger operator $H=-\frac 1 2
\partial^2_x+a\,x$ with $x\in\mathbb{R}^+$ and $a$ a real constant.
We can compute the contribution to $A_{7\over 2}$
proportional to $(\partial_y V)^2$ from the order $\lambda^{-4}$ term of the
asymptotic expansion, 
since the complete $A_{7\over 2}$ for the linear potential reduces to 
\bea
A_{7\over 2}=k\ \sqrt{\frac{\pi}{8}}\ \frac{(\partial_y V)^2}{2^6} 
\label{Ak}
\eea
with $(\partial_y V)^2=a^2$ for a constant $k$ which has to be fixed.

From the eigenfunctions of $H$ one obtains a simple expression for the kernel
$G(x,x',\lambda)$ of the resolvent $(H+\lambda)^{-1}$
\begin{equation}
    G(x,x',\lambda)=-\frac{2}{W}\left\{\theta(x'-x)L(x)R(x')+
    \theta(x-x')L(x')R(x)\right\}
\end{equation}
where $R(x)$ can be written in terms of the modified Bessel
function $K_\nu(z)$
\begin{equation}
    R(x)=\sqrt{ax+\lambda}\ K_{1/3}
    \left(\frac{[2(ax+\lambda)]^{3/2}}{3a}\right)
\end{equation}
and $L(x)$ is given by
\begin{equation}
    \begin{array}{c}
    L(x)=\sqrt{ax+\lambda}\left\{
    K_{1/3}\left(\frac{(2\lambda)^{3/2}}{3a}\right)
    I_{{1}/{3}}\left(\frac{[2(ax+\lambda)]^{3/2}}{3a}\right)-\right.\\ \\
    \left.\left.\mbox{}-I_{1/3}\left(\frac{(2\lambda)^{3/2}}{3a}\right)
    K_{1/3}\left(\frac{[2(ax+\lambda)]^{3/2}}{3a}\right)\right]
    \right\}\end{array}
\end{equation}
for Dirichlet boundary conditions at $x=0$, and
\begin{equation}
    \begin{array}{c}
    L(x)=\sqrt{ax+\lambda}\left\{\sqrt{2}\lambda
    \left[K'_{1/3}\left(\frac{(2\lambda)^{3/2}}{3a}\right)
    I_{1/3}\left(\frac{[2(ax+\lambda)]^{3/2}}{3a}\right)-\right.\right.\\ \\
    \left.\mbox{}-I'_{1/3}\left(\frac{(2\lambda)^{3/2}}{3a}\right)
    K_{1/3}\left(\frac{[2(ax+\lambda)]^{3/2}}{3a}\right)\right]+\\
    \\
    +\frac{a}{2\sqrt{\lambda}}
    \left[K_{1/3}\left(\frac{(2\lambda)^{3/2}}{3a}\right)
    I_{{1}/{3}}\left(\frac{[2(ax+\lambda)]^{3/2}}{3a}\right)-\right.\\ \\
    \left.\left.\mbox{}-I_{1/3}\left(\frac{(2\lambda)^{3/2}}{3a}\right)
    K_{1/3}\left(\frac{[2(ax+\lambda)]^{3/2}}{3a}\right)\right]
    \right\}
    \end{array}
\end{equation}
for Neumann boundary conditions at $x=0$. $W$ is the Wronskian
between the corresponding $L(x)$ and $R(x)$.

From these expressions one can obtain the bulk and boundary
contributions to the asymptotic expansion of the resolvent trace
\begin{equation}
    {\rm Tr}\ (H+\lambda)^{-1}=\int_0^\infty dx\ G(x,x,\lambda)~.
\end{equation}
From the asymptotic behavior of the modified Bessel
functions for large values of their arguments, it is
simple to see that the boundary contributions of order
$\lambda^{-4}$ and proportional to $a^2$ are given by
   $ -\frac{15}{128}a^2\,\lambda^{-4}$
for Dirichlet boundary conditions, and
   $ \frac{21}{128}a^2\,\lambda^{-4}$
for Neumann boundary conditions. Comparing these values with
 (\ref{asyres}) and (\ref{Ak}), we obtain $k=-5$ for Dirichlet boundary conditions, and
$k=7$ for Neumann boundary conditions. These are the explicit numerical values
of some multiple integrals, produced by our general method, 
that we were not able to evaluate directly.

\section{Conclusions}
We have developed a path integral method to
generalize the worldline formalism to manifolds with boundaries.  
We considered the flat manifold ${\mathbb R}_+ \times {\mathbb R}^{D-1}$ and
made use of an image 
charge to extend the problem to $\mathbb{R}^{D}$. 
Smooth potentials on $\mathbb{R}_+ \times {\mathbb R}^{D-1}$ 
extend to potentials on ${\mathbb R}^D$  which 
are generically not smooth at $y=0$.
Our proposal to deal with this situation is in principle  applicable 
to any perturbative order, but the explicit calculations become 
increasingly harder as the perturbative order increases.
It could be useful to develop simpler strategies to evaluate
the path integral in the presence of non-smooth potentials.
This would make it easier to extend the method to curved manifolds 
with boundaries. Nevertheless we have been able to show 
the feasibility of the worldline formalism 
in the presence of boundaries, and used it to calculate two new
coefficients, $A_3$ and $A_\frac{7}{2}$, for a flat space hamiltonian
with a generic scalar potential. On top of using curved manifolds, one could 
try to extend the worldline formalism on spaces with boundaries to include 
fields with spin and use more generic boundary conditions. 

\acknowledgments{OC would like to thank the organizers of the
Simons Workshop in Mathematics and Physics 2006, held at SUNY at Stony Brook,
for hospitality and partial support while parts of this work were
completed. PAGP acknowledges financial support from CONICET and partial
support from UNLP (Subsidio a J\'ovenes Investigadores and Proj. 11/X381) and from
Coimbra Group. PAGP would also like to thank the Dipartimento di Fisica
dell'Universit\`a di Bologna for friendly hospitality. } 


\end{document}